# BAGNet: Bidirectional Aware Guidance Network for Malignant Breast lesions Segmentation

Gongping Chen
*Tianjin Key Laboratory of Intelligent Robotics, College of Artificial Intelligence*
Nankai *University*
Tianjin, China
cgp110@stu.xhu.edu.cn

Yuming Liu
*Tianjin Key Laboratory of Intelligent Robotics, College of Artificial Intelligence*
Nankai *University*
Tianjin, China
1711474@mail.nankai.edu.cn

Yu Dai
*Tianjin Key Laboratory of Intelligent Robotics, College of Artificial Intelligence*
Nankai *University*
Tianjin, China
daiyu@nankai.edu.cn

Jianxun Zhang
*Tianjin Key Laboratory of Intelligent Robotics, College of Artificial Intelligence*
Nankai *University*
Tianjin, China
zhangjx@nankai.edu.cn

Liang Cui
*Department of Urology
Civil Aviation General Hospital*
Beijing, China
m15225042110@163.com

Xiaotao Yin
*Department of Urology
Fourth Medical Center of
Chinese PLA General Hospital*
Beijing, China
yxtfwy@163.com

*Abstract*—Breast lesions segmentation is an important step of computer-aided diagnosis system, and it has attracted much attention. However, accurate segmentation of malignant breast lesions is a challenging task due to the effects of heterogeneous structure and similar intensity distributions. In this paper, a novel bidirectional aware guidance network (BAGNet) is proposed to segment the malignant lesion from breast ultrasound images. Specifically, the bidirectional aware guidance network is used to capture the context between global (low-level) and local (high-level) features from the input coarse saliency map. The introduction of the global feature map can reduce the interference of surrounding tissue (background) on the lesion regions. To evaluate the segmentation performance of the network, we compared with several state-of-the-art medical image segmentation methods on the public breast ultrasound dataset using six commonly used evaluation metrics. Extensive experimental results indicate that our method achieves the most competitive segmentation results on malignant breast ultrasound images.

*Keywords*—Breast ultrasound, Automatic segmentation, Deep learning, Bidirectional guidance.

## I. Introduction

Breast cancer is a terrible disease that seriously threatens the health of women. Regular screening is essential for the prevention and diagnosis of breast cancer, due to the characteristics of concealed and multiple pathogenic factors. Currently, breast ultrasound (BUS) imaging is a widely used clinical screening method due to its advantages of painless, high sensitivity, noninvasive, and low cost [1]. BUS image segmentation can help characterize tissues and improve diagnosis, and it is an important part of BUS computer-aided diagnosis (CAD) systems [2], [3]. However, the interference of various factors (such as speckle noise, blurred boundaries heterostructure and irregular shape) make accurate BUS image segmentation a challenging task. Especially for malignant breast ultrasound image segmentation, see Fig. 1 for more details.

This work is supported by the National Natural Science Foundation of China (grant number: U1913207, 51875394). Corresponding author: Yu Dai

XXX-X-XXXX-XXXX-X/XX/$XX.00 ©20XX IEEE

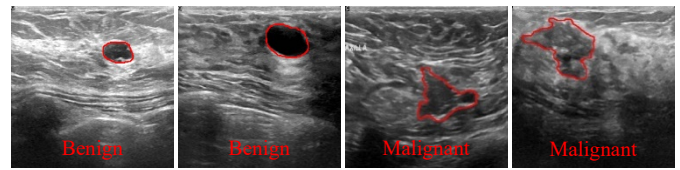

Fig. 1. From left to right are benign and malignant breast ultrasound images. The red curve is the boundary of the lesion. Compared with benign lesions, malignant lesions have an irregular shape and a severe cascade with surrounding tissue in ultrasound images. In addition, the boundary of malignant lesions is insignificant and the intensity distribution is closer to the surrounding tissue (background).

At present, many segmentation methods have been developed to segment the breast lesion accurately from ultrasound images [4]. These segmentation approaches can be classified into three types: manual, semi-automatic and automatic according to the degree of human intervention in the segmentation process [5]. Previously, the contours of breast lesions were manually annotated by radiologists. Manually segmenting lesions from ultrasound images is time-consuming and error-prone [6]. To alleviate the challenges of manual segmentation, many semi-automatic methods have been developed to segment BUS images [5]. Compared with manual annotation, these semi-automatic segmentation methods not only reduce the variance of manual segmentation of breast lesions, but also further improve work efficiency. However, Yin et al. [7] pointed out that many semi-automatic methods rely on hand-crafted features. Therefore, it is very meaningful to segment the lesion automatically and reliably from BUS images. Recently, convolutional neural networks (CNNs) have been widely used in medical image segmentation and have shown significant advantages compared to previous segmentation methods [8]–[13]. Among them, FCNN [14] and U-net [9] are typical CNN models, which have received extensive attention in BUS images segmentation based on their core architectures [15]–[17]. However, the surrounding tissue similar to the lesion affected the segmentation performance of the network, as shown in Fig. 1. To better utilize the contextual information of images, many methods introduce dilated convolution and pooling operations to obtain larger receptive fields [18]–[20]. However, these operations cannot capture contextual information from the global view and only consider dependencies on the spatial domain. Recently, capturing long-range and global dependency information has been shown to improve segmentation accuracy [21], [22]. Although these methods improve the segmentation accuracy of lesion regions, learning limited long-range and global features in deeper convolutional layers will affect the performance of the segmentation network [23]. How to fully exploit the global and long-range dependencies to improve the segmentation accuracy of medical images remains a challenging task. Most importantly, existing methods tend to address a general problem while ignoring the differences between benign and malignant lesions. Accurately segmenting malignant lesions is more difficult than segmenting benign lesions, as shown in Fig. 1.

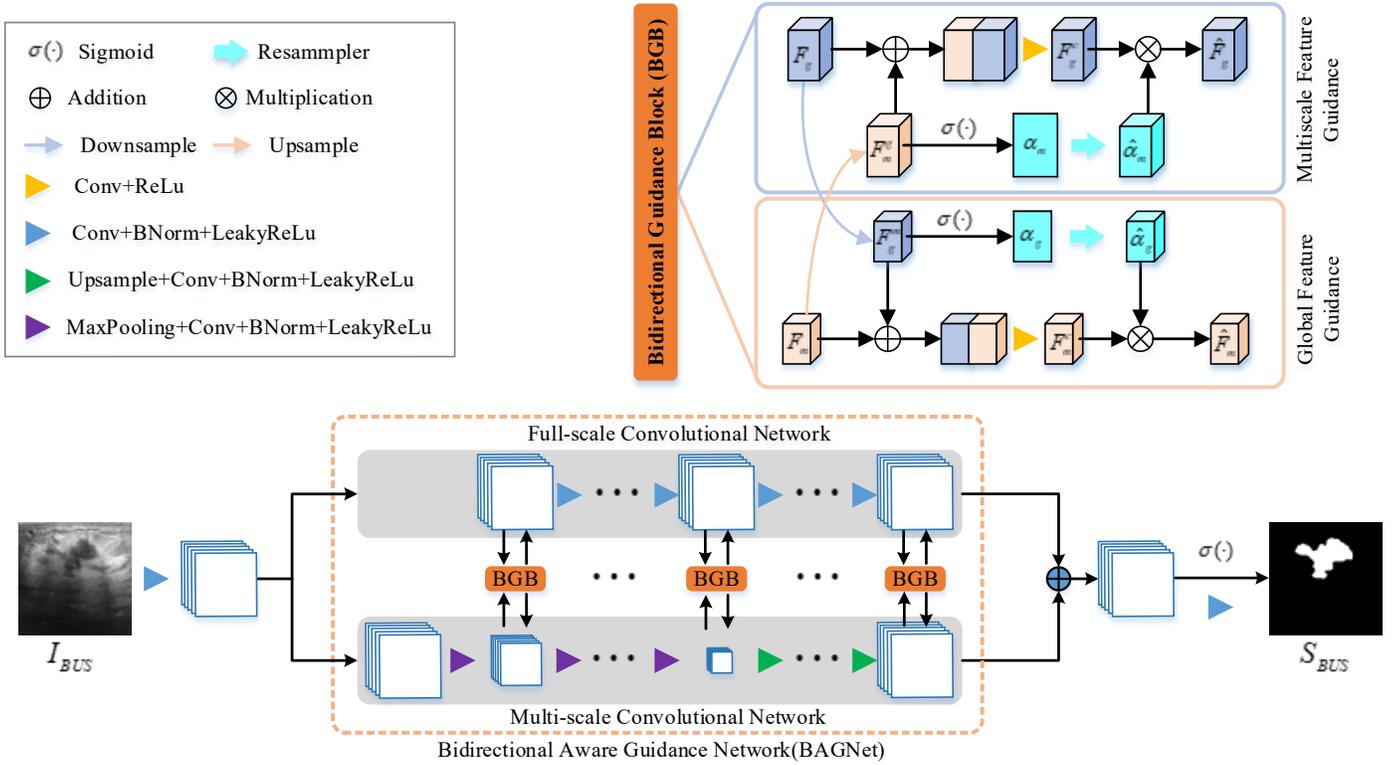

Fig. 2. The overall architecture of the segmentation network. The bidirectional aware-guided network (BAGNet) is mainly composed of two branches, one branch is responsible for extracting global information, and the other branch is responsible for capturing multi-scale feature information.

To address the above challenges, a novel bidirectional aware guidance network (BAGNet) is developed to segment malignant lesions from the BUS image, as shown in Fig. 2. The design of bidirectional aware guidance network can capture the context between global (low-level) and local (high-level) features, which are easily lost in deep convolutional layers. In the segmentation process, the introduction of the global perspective features can reduce the interference of surrounding tissues (background) on the lesion area and improve the segmentation accuracy of the network.

## II. METHOD

### A. Overview

Fig. 2 illustrates the overall architecture of the bidirectional aware guidance network (denoted as BAGNet). Our ultimate goal is to use the network to predict the mask of malignant breast lesions from BUS images. Specifically, we first utilize the convolution operation to generate a set of coarse feature maps. Then, the BAGNet constrains the coarse feature maps from the global view. Finally, the feature map generated by BAGNet is subjected to the convolution operation and a sigmoid operation to obtain the final segmentation result.

### B. Bidirectional Aware Guidance Network (BAGNet)

As shown in Fig. 2, BAGNet is mainly composed of parallel full-scale convolutional network and multi-scale convolutional network, which are responsible for the extraction of global features and multi-scale features respectively. The number of convolutional layers included in full-scale convolutional network and multi-scale convolutional network are 8 and 9, respectively, and each convolutional layer with kernel-size is $3\times3$ followed by a batch normalization layer and a linear activation layer. The number of convolution kernels in full-scale convolutional network and multi-scale convolutional network are 32 and 64, respectively. Different from full-scale convolutional network, multi-scale convolutional network also includes four down-sampling operations and four up-sampling operations. In order to realize the bidirectional calibration of global features and local features, we design a bidirectional guidance block (BGB). Eight BGBs are included in BAGNet.

As shown in Fig. 2, the BGB block contains two parts: global feature guidance and multiscale feature guidance. For the global feature guidance branch, we first down-sample the input global feature map $F_g$ to the same spatial dimension as the multi-scale feature map $F_m$, and the result is denoted as:

$$F_g^m = S_D(F_g) \quad (1)$$

where $S_D(\cdot)$ denotes the down-sampling operation. Subsequently, $F_g^m$ and $F_m$ are merged and fed into a convolutional layer to obtain the feature map:

$$F_m^c = S_C(F_m \oplus F_g^m) \quad (2)$$

where $\oplus$ represents the concatenation operation, and $S_C(\cdot)$ indicates the convolution operation. At the same time, a linear projection matrix $W_g \in \mathbb{R}^{1\times1\times1}$ squeezes $F_g^m$ along the channel direction into one channel, and then a sigmoid activation $\sigma(\cdot)$ is used to obtain the global attention map:

$$\alpha_g = \sigma(W_g \cdot F_g^m) \quad (3)$$

where $W_g$ is implemented by a convolution, $\alpha_g \in [0,1]$ is the spatial attention map for $F_m^c$. To calibrate the feature $F_m^c$, $\alpha_g$ is resampled to obtain a new attention map $\hat{\alpha}_g$ with the same number of channels as $F_m^c$. The feature map after calibration by $\hat{\alpha}_g$ can be expressed as:

$$\hat{F}_m = \hat{\alpha}_g \otimes F_m^c \quad (4)$$

Finally, $\hat{F}_m$ is input to the next convolution stage.

Similarly, the operations performed in the multiscale feature guidance branch are similar to the global feature guidance branch. we first up-sample the input multiscale feature map $F_m$ to the same spatial dimension as the global feature map $F_g$, and the result is denoted as:

$$F_m^g = S_U(F_m) \quad (5)$$

where $S_U(\cdot)$ denotes the up-sampling operation. Subsequently, $F_m^g$ and $F_g$ are merged and fed into a convolutional layer to obtain the feature map:

$$F_g^c = S_C(F_g \oplus F_m^g) \quad (6)$$

where $\oplus$ represents the concatenation operation, and $S_C(\cdot)$ indicates the convolution operation. At the same time, a linear projection matrix $W_m \in \mathbb{R}^{1\times1\times1}$ squeezes $F_m^g$ along the channel direction into one channel, and then a sigmoid activation $\sigma(\cdot)$ is used to obtain the global attention map:

$$\alpha_m = \sigma(W_m \cdot F_m^g) \quad (7)$$

where $W_m$ is implemented by a convolution, $\alpha_m \in [0,1]$ is the spatial attention map for $F_g^c$. To calibrate the feature $F_g^c$, $\alpha_m$ is resampled to obtain a new attention map $\hat{\alpha}_m$ with the same number of channels as $F_g^c$. The feature map after calibration by $\hat{\alpha}_m$ can be expressed as:

$$\hat{F}_g = \hat{\alpha}_m \otimes F_g^c \quad (8)$$

Finally, $\hat{F}_g$ is input to the next convolution stage. Supervision of the ground-truth mask is added at the end of BAGNet to guide the segmentation process of the network.

### C. Experimental Parameters

In this paper, we adopted binary-cross entropy (BCE) as loss function. The Adam optimizer was chosen to train our network with an initial learning rate of 0.001. After multiple cross-validations, the epoch size and batch size were set to 50 and 12, respectively.

### III. EXPERIMENT

### A. Datasets

We use a public benchmark BUS dataset (BUSIS) to evaluate our segmentation network. The breast tumor ultrasound dataset constructed by Al-Dhabyani et al. [24] contains a total of 780 images of 600 female patients. Among them, 133 normal cases, 437 benign cases, and 210 malignant masses. The goal of this paper is to achieve accurate segmentation of malignant breast lesions, so we only use 210 malignant ultrasound images for comparative experiments. In the experiments, we use three-fold cross-validation on the malignant BUS images.

### B. Evaluation Metrics

In this paper, we use six segmentation metrics to quantitatively evaluate the segmentation performance of our method on malignant BUS images. The six segmentation metrics are Accuracy, Jaccard index, Precision, Recall, Specificity and Dice [25].

### C. Experimental Results

Six state-of-the-art medical image segmentation methods are used for comparative experimental analysis. They are U-net [9], Att U-net [26], U-net++ [11], U-net3+ [13], SegNet [27] and SATN [28]. Each segmentation method is properly trained to ensure fair comparisons. In the comparative experiment, we adopt the cross-validation method for comparison.

The evaluation scores of various segmentation methods on the malignant breast lesion are shown in Table I. Obviously, our method achieves the best performance on the segmentation of malignant lesions. The values of our method on Accuracy, Jaccard, Precision, Recall, Specificity and Dice are 92.60%, 59.71%, 75.69%, 76.99%, 96.46% and 69.93%, respectively. Compared with the second segmentation results for the malignant masses, our method improves by 0.49%, 8.78%, 12.20%, 0.97%, 2.62% and 5.65%, respectively. In addition, the advantage of the method in this paper is more obvious compared with the third-ranked result. The segmentation results of U-net on malignant masses show that the segmentation performance of this method on BUS image is not ideal. By comparing the segmentation results of the U-net variant network, it can be concluded that the segmentation performance of the network can be improved by additional or more complex operations. It is worth noting that SegNet achieves relatively competitive results on the segmentation of breast lesions. This indicates that the learning of location information is beneficial to the improvement of segmentation performance. STAN achieves good results on the segmentation of small breast tumors, but not ideal on the segmentation of malignant breast lesions.

TABLE I. COMPARISON WITH DIFFERENT METHODS ON MALIGNANT BREAST LESIONS (MEAN±STD).

| Method | Accuracy | Jaccard | Precision | Recall | Specificity | Dice |
|---|---|---|---|---|---|---|
| U-net | 90.85±0.78 | 51.11±2.62 | 64.96±2.55 | 68.86±4.27 | 93.63±1.28 | 63.47±2.38 |
| STAN | 91.04±0.76 | 51.11±2.38 | 64.74±3.72 | 70.96±6.75 | 93.69±1.56 | 62.60±1.97 |
| Att U-net | 91.22±0.91 | 51.12±2.35 | 61.62±0.97 | 72.57±2.17 | 93.12±1.00 | 62.95±2.14 |
| U-net++ | 91.53±0.84 | 54.03±3.03 | 65.50±2.94 | 73.43±2.10 | 93.73±1.31 | 65.52±2.75 |
| U-net3+ | 91.63±0.78 | 54.77±3.55 | 65.78±2.66 | 74.38±3.21 | 93.82±1.06 | 66.19±3.37 |
| SegNet | 92.15±0.64 | 54.89±1.78 | 63.79±2.65 | 76.25±4.02 | 94.00±1.14 | 65.90±1.97 |
| **Ours** | **92.60±0.77** | **59.71±3.75** | **75.69±2.51** | **76.99±3.58** | **96.46±0.85** | **69.93±3.63** |

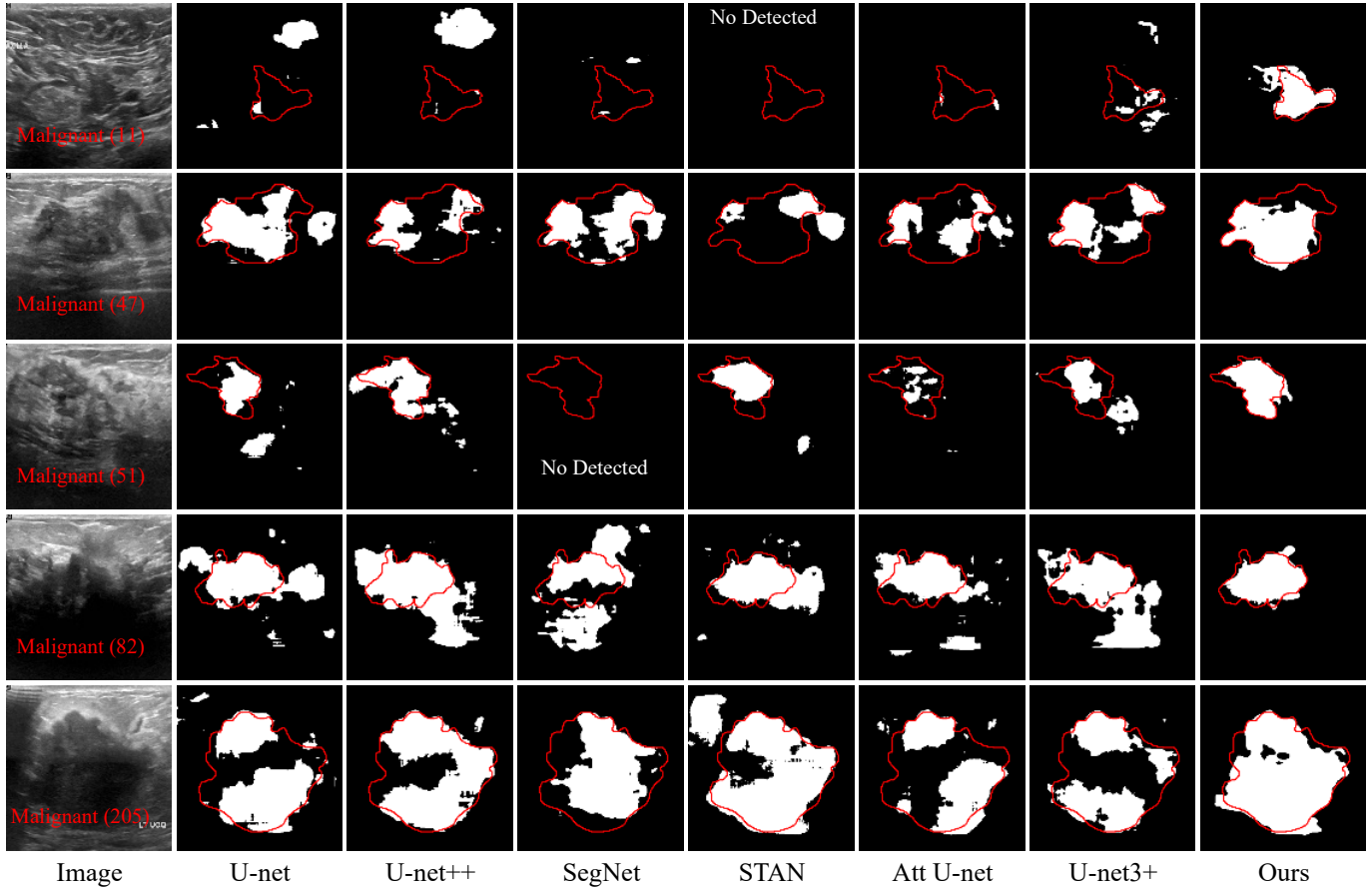

Fig. 3. Segmentation results of malignant breast lesions by different methods. The red curve is the boundary of the malignant breast lesion.

Fig. 3 shows the segmentation results of different segmentation methods on malignant BUS images. Apparently, our method has the best segmentation performance on BUS images and achieves segmentation results closer to the ground-truth masks. Compared with other segmentation methods, our method can effectively reduce the missed detection rate and false detection rate of breast lesions segmentation. This suggests that global feature guidance can help improve the segmentation accuracy of malignant breast lesions. As shown in Fig. 3, the surrounding tissue in the ultrasound image with similar intensity distribution of the lesion seriously affects the segmentation accuracy of the lesion. To make matters worse, some network architectures fail to detect the lesion region from the BUS images, as shown in the first, and third rows. Fortunately, our method is able to alleviate the perturbation of similar surrounding tissues (background) and has better segmentation performance. According to the segmentation results of the comparison methods, they have poor segmentation performance on breast lesions segmentation with serious false and missed detections. Comprehensive analysis can conclude that our method is more robust and more suitable for malignant breast lesions segmentation.

## IV. CONCLUSION

This paper presented a novel bidirectional aware guidance network (BAGNet) to segment malignant lesions from BUS images. The design of the bidirectional aware guidance network captures the context between global (low-level) and local (high-level) features from the saliency map, which can reduce the interference of intensity-similar surrounding tissues (background) on malignant lesions segmentation. Experimental results show that our method achieves the most competitive results on the malignant breast lesions segmentation.

## REFERENCES

[1] J. A. Noble and D. Boukerroui, "Ultrasound image segmentation: a survey," *IEEE Trans Med Imaging*, vol. 25, no. 8, pp. 987–1010, 2006, doi: 10.1109/tmi.2006.877092.

[2] Q. Huang, Y. Luo, and Q. Zhang, "Breast ultrasound image segmentation: a survey," *Int. J. Comput. Assist. Radiol. Surg.*, vol. 12, no. 3, pp. 493–507, 2017.

[3] Y. Luo, Q. Huang, and X. Li, "Segmentation information with attention integration for classification of breast tumor in ultrasound image," *Pattern Recognit.*, vol. 124, p. 108427, 2022, doi: 10.1016/j.patcog.2021.108427.

[4] J. Bai, R. Posner, T. Wang, C. Yang, and S. Nabavi, "Applying deep learning in digital breast tomosynthesis for automatic breast cancer detection: A review," *Med. Image Anal.*, vol. 71, p. 102049, 2021.

[5] M. Xian, Y. Zhang, H.-D. Cheng, F. Xu, B. Zhang, and J. Ding, "Automatic breast ultrasound image segmentation: A survey," *Pattern Recognit.*, vol. 79, pp. 340–355, 2018.

[6] S. Liu *et al.*, "Deep Learning in Medical Ultrasound Analysis: A Review," *Engineering*, vol. 5, no. 2, pp. 261–275, 2019, doi: 10.1016/j.eng.2018.11.020.

[7] S. Yin *et al.*, "Automatic kidney segmentation in ultrasound images using subsequent boundary distance regression and pixelwise classification networks," *Med Image Anal*, vol. 60, p. 101602, 2020, doi: 10.1016/j.media.2019.101602.

[8] D. Dayakshini, S. Kamath, K. Prasad, and K. V. Rajagopal, "Segmentation of Breast Thermogram Images for the Detection of Breast Cancer – A Projection Profile Approach," *J. Image Graph.*, vol. 3, no. 1, pp. 47–51, 2015, doi: 10.18178/joig.3.1.47-51.

[9] O. Ronneberger, P. Fischer, and T. Brox, "U-net: Convolutional networks for biomedical image segmentation," in *International Conference on Medical image computing and computer-assisted intervention*, 2015, pp. 234–241.

[10] N. Borzooie, H. Danyali, and M. S. Helfroush, "Modified Density-Based Data Clustering for Interactive Liver Segmentation," *J. Image Graph.*, vol. 6, no. 1, pp. 84–87, 2018, doi: 10.18178/joig.6.1.84-87.

[11] Z. Zhou, M. Siddiquee, N. Tajbakhsh, and J. Liang, "UNet++: Redesigning Skip Connections to Exploit Multiscale Features in Image Segmentation," *IEEE Trans. Med. Imaging*, vol. 39, no. 6, pp. 1856–1867, 2020.

[12] A. O. Joshua, F. V. Nelwamondo, and G. Mabuza-Hocquet, "Blood Vessel Segmentation from Fundus Images Using Modified U-net Convolutional Neural Network," *J. Image Graph.*, vol. 8, no. 1, pp. 21–25, 2020, doi: 10.18178/joig.8.1.21-25.

[13] H. Huang, L. Lin, R. Tong, H. Hu, and J. B. T.-I. 2020-2020 I. I. C. on A. Wu Speech and Signal Processing (ICASSP), "UNet 3+: A Full-Scale Connected UNet for Medical Image Segmentation," 2020.

[14] J. Long, E. Shelhamer, and T. Darrell, "Fully convolutional networks for semantic segmentation," in *Proceedings of the IEEE conference on computer vision and pattern recognition*, 2015, pp. 3431–3440.

[15] M. H. Yap *et al.*, "Automated Breast Ultrasound Lesions Detection Using Convolutional Neural Networks," *IEEE J. Biomed. Heal. Informatics*, vol. 22, no. 4, pp. 1218–1226, 2018, doi: 10.1109/JBHI.2017.2731873.

[16] R. Almajalid, J. Shan, Y. Du, and M. Zhang, "Development of a deep-learning-based method for breast ultrasound image segmentation," in *2018 17th IEEE International Conference on Machine Learning and Applications (ICMLA)*, 2018, pp. 1103–1108.

[17] K. Huang, Y. Zhang, H.-D. Cheng, P. Xing, and B. Zhang, "Fuzzy semantic segmentation of breast ultrasound image with breast anatomy constraints," *arXiv Prepr. arXiv1909.06645*, 2019.

[18] Z. Zhuang, N. Li, A. N. Joseph Raj, V. G. V Mahesh, and S. Qiu, "An RDAU-NET model for lesion segmentation in breast ultrasound images," *PLoS One*, vol. 14, no. 8, p. e0221535, 2019.

[19] Y. Hu *et al.*, "Automatic tumor segmentation in breast ultrasound images using a dilated fully convolutional network combined with an active contour model," *Med. Phys.*, vol. 46, no. 1, pp. 215–228, 2019.

[20] H. Lee, J. Park, and J. Y. Hwang, "Channel attention module with multiscale grid average pooling for breast cancer segmentation in an ultrasound image," *IEEE Trans. Ultrason. Ferroelectr. Freq. Control*, vol. 67, no. 7, pp. 1344–1353, 2020.

[21] T. Dou, L. Zhang, H. Zheng, and W. Zhou, "Local and non-local deep feature fusion for malignancy characterization of hepatocellular carcinoma," in *International Conference on Medical Image Computing and Computer-Assisted Intervention*, 2018, pp. 472–479.

[22] K. Qi *et al.*, "X-net: Brain stroke lesion segmentation based on depthwise separable convolution and long-range dependencies," in *International conference on medical image computing and computer-assisted intervention*, 2019, pp. 247–255.

[23] C. Xue *et al.*, "Global guidance network for breast lesion segmentation in ultrasound images," *Med. Image Anal.*, vol. 70, p. 101989, 2021.

[24] W. Al-Dhabyani, M. Gomaa, H. Khaled, and A. Fahmy, "Dataset of breast ultrasound images," *Data Br.*, vol. 28, p. 104863, 2020.

[25] G. Chen, J. Yin, Y. Dai, J. Zhang, X. Yin, and L. Cui, "A novel convolutional neural network for kidney ultrasound image segmentation," *Comput. Methods Programs Biomed.*, vol. 218, p. 106712, 2022, doi: 10.1016/j.cmpb.2022.106712.

[26] O. Oktay *et al.*, "Attention U-Net: Learning Where to Look for the Pancreas," no. Midl, 2018, [Online]. Available: http://arxiv.org/abs/1804.03999.

[27] V. Badrinarayanan, A. Kendall, and R. Cipolla, "SegNet: A Deep Convolutional Encoder-Decoder Architecture for Image Segmentation," *IEEE Trans Pattern Anal Mach Intell*, vol. 39, no. 12, pp. 2481–2495, 2017, doi: 10.1109/TPAMI.2016.2644615.

[28] B. Shareef, M. Xian, and A. Vakanski, "Stan: Small tumor-aware network for breast ultrasound image segmentation," in *2020 IEEE 17th International Symposium on Biomedical Imaging (ISBI)*, 2020, pp. 1–5.

**Authors' background**

| Name | Prefix | Research Field | Email | Personal website |
|---|---|---|---|---|
| Gongping Chen | PhD Candidate | Medical image analysis, Deep Learning | cgp110@stu.xhu.edu.cn | https://github.com/CGPzy |
| Yu Dai | Full Professor | Medical surgical robot | daiyu@nankai.edu.cn | https://ai.nankai.edu.cn/info/1033/4187.htm |
| Jianxun Zhang | Full Professor | Medical surgical robot | zhangjx@nankai.edu.cn | https://ai.nankai.edu.cn/info/1033/2787.htm |
| Yuming Liu | Master Student | Image segmentation, Deep learning | 1711474@mail.nankai.edu.cn | No |
| Liang Cui | Full Professor | Minimally invasive surgical robot | m15225042110@163.com | https://www.mhzyy.cn/Detail/index.html?id=53&aid=4361 |
| Xiaotao Yin | Full Professor | Minimally invasive surgical robot | yxtfwy@163.com | No |